\documentclass[12pt]{article}
\usepackage{amssymb}

\usepackage{graphicx}
\usepackage{amsmath}

\input{tcilatex}

\begin{document}

\title{{\LARGE THE EQUIVALENCE PRINCIPLE AS A SYMMETRY}}
\date{}
\author{}
\maketitle

\begin{center}
\bigskip Paul S. Wesson$^{1,\,2}$
\end{center}

\begin{enumerate}
\item \bigskip Dept. Physics, University of Waterloo, Waterloo, Ontario N2L
3G1, Canada \linebreak Keywords: General relativity, Induced-matter theory,
brane theory

Pacs: 04.50+h, 04.20.Cv, 11.10.Kk

\item Correspondence: mail to (1) above, fax to (519) 746-8115,

email to wesson@astro.uwaterloo.ca. \pagebreak
\end{enumerate}

\begin{center}
{\LARGE THE EQUIVALENCE PRINCIPLE AS A SYMMETRY}
\end{center}

\underline{{\LARGE Abstract}}

It is shown that the extra coordinate of 5D induced-matter and membrane
theory is related in certain gauges to the inertial rest mass of a test
particle. \ This implies that the Weak Equivalence Principle is a geometric
symmetry, valid only in the limit in which the test mass is negligible
compared to the source mass. \ Exact solutions illustrate this, and show the
way to possible resolutions of the cosmological-constant and hierarchy
problems.

\section{\protect\underline{Introduction}}

\ \ \ \ \ The Weak Equivalence Principle is commonly taken to mean that in a
gravitational field the acceleration of a test particle is independent of
the properties of the latter, including its rest mass. \ Recently, however,
the extension of 4D general relativity to 5D has led to the isolation of a
fifth force, which exists for both induced-matter theory [1, 2] and membrane
theory [3, 4]. \ These two versions of what used to be called Kaluza-Klein
theory allow dependence on an extra coordinate $l$, and it is now known that
their field equations are essentially the same [5]. \ In both theories, the
extra force per unit rest mass is an acceleration which is inertial in the
Einstein sense, arising from the motion in the fifth dimension with respect
to the 4D part of the manifold which we call spacetime [6]. \ This extra
acceleration has already been related to the (inertial) rest mass $m$ of a
test particle [1, 3] in certain choices of coordinate frame (or gauge), and
in general its presence represents a technical violation of the 4D WEP. \
Such violations of the 4D WEP in $N\left( >4\right) D$ field theory have
been mentioned before [7-11; for a short review see ref. 6, pp. 85-88]. \
However, the WEP is known from experiments conducted from the time of
Galileo to now to be obeyed with an accuracy of at least 1 part in 10$^{11}$
[12]. \ The purpose of the present work is to clarify the status of the 4D
Weak Equivalence Principle in $N\left( >4\right) D$ extensions of general
relativity. \ We will do this for 5D; but the extension to higher $N$ as in
10D superstrings, 11D supergravity and 26D string theory is
straight-forward, and in fact guaranteed by Campbell's theorem [13-15]. \
The plan is as follows: (a) Marshall extant mathematical results [16-22],
showing that they have the consistent physical interpretation that the extra
coordinate $l$ measures the (inertial) rest mass of a test particle $m$; (b)
Illustrate the cogency of this inference by giving 3 exact $l$-dependent
solutions of the 5D field equations which generalize the 4D de Sitter
solution of general relativity as widely used in particle physics [23-27],
thereby generalizing the concept of ``\underline{the} vacuum'' and opening a
way to a resolution of the cosmological-constant problem; (c) Use the scalar
potential as a classical analog of the Higgs field [6, 28], leading to an
expression for the masses of real particles which avoids the hierarchy
problem; (d) Conclude that the WEP is a geometric symmetry, valid only in
the limit where the mass of a test particle is negligible compared to the
mass of the source, thus supporting new endeavors [29, 30] to look for
violations.

\section{\protect\underline{The Nature of the Fifth Coordinate}}

\ \ \ \ There are 5 degrees of coordinate freedom in an unrestricted 5D
Riemannian manifold, of which 4 can be used to remove the potentials of
electromagnetic type, giving the line element $dS^{2}=g_{AB}dx^{A}dx^{B}=g_{%
\alpha \beta }\left( x^{\gamma },\;l\right) dx^{\alpha }dx^{\beta }+\epsilon
\Phi ^{2}dl^{2}\left( A=O,\,123,\,4;\,\alpha =O,\,123\right) $. The
signature is $+\left( ---\right) \epsilon $ where $\epsilon =\pm 1$ is not
restricted by Campbell's theorem [13-15], the usual $\epsilon =-1$ admitting
particle-like solutions and $\epsilon =+1$ admitting wave-like solutions
[22]. \ The coordinates are $x^{0}=t$ for time, $x^{123}=xyz$ (or $r\theta
\phi $) for space and an extra one $x^{4}=l$. \ All will be taken to have
physical dimensions of length, and the constants $c,\,G,\,h$ will usually be
absorbed by a choice of units. \ It will turn out to be useful to defer
usage of the fifth degree of coordinate freedom, though in principle it is
available to suppress the scalar potential $\left( \Phi \right) $ or to
restrict the velocity in the fifth dimension. \ With regard to velocities,
we wish to make contact with 4D physics couched in terms of $%
ds^{2}=g_{\alpha \beta }\,dx^{\alpha }dx^{\beta }$ and 4-velocites $%
u^{\alpha }\equiv dx^{\alpha }\diagup ds$. \ We will therefore parametize
motions in terms of the elements of 4D proper time $ds$, a choice which also
allows us to handle null 5D paths with $dS=0$ [2, 4]. \ With this setup, we
can make several observations on the physical nature of the fifth coordinate.

(i) The extra force which appears when the manifold is extended from 4D to
5D has been derived in different ways for induced-matter theory [1, 2] and
brane theory [3, 4]. \ But a generic and shorter way is as follows. \ The
relation%
\begin{equation}
g_{\alpha \beta }\left( x^{\gamma },\;l\right) u^{\alpha }u^{\beta }=1
\end{equation}%
is a normalization condition on the 4-velocities. \ When multiplied by the
inertial rest mass $m$ of a test particle, it gives the usual relation $%
E^{2}-p^{2}=m^{2}$ where $E$ is the energy and $p$ is the 3-momentum. \
(Alternatively, $p^{\alpha }p_{\alpha }=m^{2}$ where $p^{\alpha }\equiv
mu^{\alpha }$ are the 4-momenta.) \ There is actually no information in (1)
about the possibility that $m=m\left( s\right) $, which applies for example
to the case of a rocket which loses mass as it burns fuel and so
accelerates. \ The acceleration in such a case is given by appeal to the law
of conservation of linear momentum (see below). \ However, we can consider
the effect of a slight change in the 5D coordinates (including $l$) by
differentiating (1) with respect to $s$. \ Doing this and using symmetries
under the exchange of $\alpha $ and $\beta $ to introduce the Christoffel
symbols $\Gamma _{\alpha \beta }^{\mu }$, there comes 
\begin{equation}
2g_{\alpha \mu }u^{\alpha }\left( \frac{du^{\mu }}{ds}+\Gamma _{\beta \gamma
}^{\mu }u^{\beta }u^{\gamma }\right) +\frac{\partial g_{\alpha \beta }}{%
\partial l}\frac{dl}{ds}u^{\alpha }u^{\beta }=0\;\;\;.
\end{equation}%
This reveals that in addition to its usual 4D geodesic motion (the part
inside the parenthesis), a particle feels a new acceleration (or force per
unit mass). \ It is due to the motion of the 4D frame with respect to the
fifth dimension, and is \underline{parallel} to the 4-velocity $u^{\mu }$. \
Explicitly, the parallel acceleration is 
\begin{equation}
P^{\mu }=-\frac{1}{2}\left( \frac{\partial g_{\alpha \beta }}{\partial l}%
u^{\alpha }u^{\beta }\right) \frac{dl}{ds}u^{\mu }\;\;\;\;_{.}
\end{equation}%
This has no analog in 4D field theory, including Einstein gravity and
Maxwell electromagnetism, where forces are orthogonal to the velocities and
obey $F^{\mu }u_{\mu }=0.$ \ (Another way of seeing that an extra force must
appear in the extension from 4D to 5D is to note as in ref. 1 that if $%
F^{A}u_{A}=0$ then $F^{\mu }u_{\mu }=-F^{4}u_{4}\neq 0$.) \ To investigate
(3), we can evaluate it in the canonical coordinate system, which is so
called because it leads to great algebraic simplification of the geodesic
equation and the field equations (see below) and has been extensively used
[6, 17-19, 22]. \ Then $g_{\alpha \beta }\left( x^{\gamma },\,l\right)
=\left( l^{2}\diagup L^{2}\right) \overline{g}_{\alpha \beta }\left(
x^{\gamma }\right) ,$ where $L$ is a length introduced for dimensional
consistency, and for vacuum 4D spacetimes is given by $L^{2}=3\diagup
\Lambda $ where $\Lambda $ is the cosmological constant [17]. \ The
acceleration (3) can now be evaluated and simplified using (1). \ Its nature
becomes clear in the Minkowski limit, when the motion of the particle is
given by%
\begin{equation}
\frac{du^{\mu }}{ds}=P^{\mu }=-\frac{1}{l}\frac{dl}{ds}u^{\mu }\;\;\;,
\end{equation}%
\begin{equation}
\text{or}\;\;\;\;\;\frac{d}{ds}\left( l\;u^{\mu }\right) =0\;\;\;\;.
\end{equation}%
The last is just the expected law of conservation of linear momentum,
provided $l=m$.

(ii) The action can be used to confirm this. \ Let us write the 5D interval
in terms of its 4D and extra parts using a coordinate system which is
perturbed from the pure canonical one noted above. \ Then with $d\overline{s}%
^{2}\equiv \overline{g}_{\alpha \beta }dx^{\alpha }dx^{\beta }$ we have%
\begin{eqnarray}
dS^{2} &=&\frac{l^{2}}{L^{2}}\overline{g}_{\alpha \beta }\left( x^{\gamma
},\;l\right) dx^{\alpha }dx^{\beta }+\epsilon \Phi \;^{2}\left( x^{\gamma
},\;l\right) dl^{2} \\
L^{2}dS^{2} &=&l^{2}d\overline{s}^{2}+\epsilon \left( \Phi L\right)
^{2}dl^{2}\;\;\;\;\;.
\end{eqnarray}%
Clearly the first term on the right-hand side here involves the conventional
element of action $md\overline{s}$ if $l=m$. \ It should be noted that even
in 4D the action should be written $\int md\overline{s}$ to account for the
possibility that the mass changes along the path, and that in 5D the
expression (6) is still general. \ So the conventional action is the 4D part
of a 5D one.

(iii) The 5D geodesic equation minimizes paths via $\delta \left( \int
dS\right) =0$, which generalizes the equations of motion in 4D and adds an
extra component for the motion in the fifth dimension. \ The working
requires the specification of a starting gauge, and is generally tedious.
(See ref. 6, pp. 132-138 and pp. 161-167 for the cases where
electromagnetism is and is not included respectively, as well as references
to other work.) \ We therefore quote here two results which are relevant. \
First, for 5D metrics which are canonical in form, the fifth force noted
above is proportional to $dl\diagup d\overline{s}$, and disappears if the
latter is zero, \underline{making the 4D part of the motion geodesic in the
usual sense}. \ It should be noted in passing that the conventional geodesic
equation is a statement about accelerations (not forces) caused by the
motion of reference frames, so this result means that 4D geodesic motion is
a special case of 5D motion, which latter is inertial in the Einstein sense.
\ Second, for 5D metrics which are independent of $x^{0}=t$, there is a
constant of the motion which is the analog of the 4D particle energy. \ When
the metric is $l$-factorized as in the canonical case, electromagnetic terms
are absent and the 3-velocity $v$ is projected out, this constant is 
\begin{equation}
E=\frac{l}{\left( 1-v^{2}\right) ^{\frac{1}{2}}}\;\;\;.
\end{equation}%
One does not have to be Einstein to see that this gives back the
conventional 4D energy provided $l$ is identified with the particle rest
mass $m$.

(iv) The field equations for 5D relativity are commonly taken in terms of
the Ricci tensor to be $R_{AB}=0\left( A=0,\;123,\;4\right) $; and by
Campbell$^{\prime }$s theorem [13-15] these contain those of general
relativity, which in terms of the Einstein tensor and the energy-momentum
tensor are $G_{\alpha \beta }=8\pi T_{\alpha \beta }$ $\left( \alpha
=0,\;123\right) $. \ Here $G_{\alpha \beta }$ is constructed as usual from
the 4D, $l$-independent parts of the 4D Ricci tensor and scalar. \ However, $%
T_{\alpha \beta }$ is an \underline{effective} or induced source,
constructed from the $l$-dependent parts of these quantities and the scalar
field $\left( g_{44}=\epsilon \Phi ^{2}\right) $. \ As such, the latter
includes parts which can be identified with conventional matter and parts
which by default refer to the ``vacuum''. \ We will return to the latter
concept below, but here we note that the general expression for the source
can be written down after some lengthy algebra. \ With the metric in the
general form $dS^{2}=g_{\alpha \beta }\left( x^{\gamma },\;l\right)
dx^{\alpha }dx^{\beta }+\epsilon \Phi ^{2}dl^{2},$ it is 
\begin{eqnarray}
8\pi T_{\alpha \beta } &=&\frac{\Phi ,_{\alpha ;\,\beta }}{\Phi }-\frac{%
\epsilon }{2\Phi ^{2}}\left\{ \frac{\Phi _{,4}\;g_{\alpha \beta ,4}}{\Phi }%
-g_{\alpha \beta ,44}+g^{\lambda \mu }g_{\alpha \lambda ,4}g_{\beta \mu
,4}\right.  \notag \\
&&\;\;\;\;\;\;\;\;\;\;\;\left. -\frac{g^{\mu \nu }g_{\mu \nu ,4}g_{\alpha
\beta ,4}}{2}+\frac{g_{\alpha \beta }}{4}\left[ g_{\;\;,4}^{\mu \nu }g_{\mu
\nu ,4}+\left( g^{\mu \nu }g_{\mu \nu ,4}\right) ^{2}\right] \right\} \;\;\;.
\end{eqnarray}%
Here a comma denotes the partial derivative with respect to $x^{4}=l$ and a
semicolon denotes the usual 4D covariant derivative. \ The expression (9) is
known to give back the conventional matter content of a wide variety of 4D
solutions [6], but in order to bolster the physical identification of $l$ we
note a generic property of it. \ For $g_{\alpha \beta ,4}=0$, (9) gives $%
8\pi T\equiv 8\pi g^{\alpha \beta }T_{\alpha \beta }=g^{\alpha \beta }\Phi
_{,\alpha ;\beta }\diagup \Phi \equiv \Phi ^{-1}\square \Phi $; but the
extra field equation $R_{44}=0$, which we will examine below, gives $\square
\Phi =0$ for $g_{\alpha \beta ,4}=0$. \ Thus $T=0$ for $g_{\alpha \beta
,4}=0 $, meaning that the equation of state is that of radiation when the
source consists of photons with zero rest mass. \ This is as expected.

(v) Algebraic arguments for $l=m$ can be understood from the physical
perspective by simple dimensional analysis. \ The latter is actually an
elementary group-theoretic technique based on the Pi theorem, and one could
argue that a complete theory of mechanics ought to use a manifold in which
spacetime is extended so as to properly take account of the \underline{three}
mechanical bases M, L, T. \ Obviously, this has to be done in a manner which
does not violate the known laws of mechanics and recognizes their use of the
three dimensional constants $c$, $G$ and $h$. \ The canonical metric of
induced-matter theory, as employed in several instances above, clearly
satisfies these criteria [1, 2]. \ But the warp metric of brane theory leads
to similar results [3, 4]; and it has indeed been argued that the two
theories are essentially the same one, expressed in different ways [5]. \
This leads to an important point: \underline{the physical identification of $%
x^{4}=l$ \ requires a choice of \ 5D \ coordinates or} \linebreak \underline{%
gauge}. \ To illustrate this, consider a 5D metric given by%
\begin{equation}
dS^{2}=\left( \frac{L}{l}\right) ^{2a}\overline{g}_{\alpha \beta }\left(
x^{\gamma }\right) dx^{\alpha }dx^{\beta }-\left( \frac{L}{l}\right)
^{4b}dl^{2}\;\;\;\;.
\end{equation}%
Here $a$, $b$ are constants which can be constrained by the full set of 5D
field equations $R_{AB}=0$ [22]. \ There are 3 choices: $a=b=0$ gives
general relativity embedded in a flat and physically innocuous extra
dimension; $a=-1,\;b=0$ gives the pure-canonical metric already discussed;
while $a=b=1$ gives a metric which looks different but is actually the
canonical one after the coordinate transformation $l\rightarrow L^{2}\diagup
l$. \ We see that the last two cases describe the same physics but in terms
of different choices of $l$. \ Temporarily introducing the relevant
constants, these are%
\begin{equation}
l_{E}=\frac{Gm}{c^{2}}\;\;,\;\;l_{P}=\frac{h}{mc}
\end{equation}%
in what may be termed the Einstein and Planck gauges. \ These represent 
\underline{convenient} choices of $x^{4}=l$, insofar as they represent
parametizations of the inertial rest mass $m$ of a test particle which fit
with known laws of 4D physics such as the conservation of momentum (see
above: the fifth force conserves $l_{E}u^{\mu }$ or $l_{P}^{-1}u^{\mu }$). \
However, 5D relativity as based on the field equations $R_{AB}=0$ is
covariant under the 5D group of transformations $x^{A}\rightarrow \overline{x%
}^{A}\left( x^{B}\right) $, which is wider than the 4D group $x^{\alpha
}\rightarrow \overline{x}^{\alpha }\left( x^{\beta }\right) $. \ Therefore
4D quantities $Q\left( x^{\alpha },\;l\right) $ will in general change under
a change of coordinates that includes $l$. \ This implies that we can only
recognize $m$ in certain gauges.

The import of the preceding comments (i)-(v) is major for the Weak
Equivalence Principle. \ In gauges like those of Einstein or Planck, or ones
close to them, the dependence of the ordinary 4D metric of spacetime on the
extra coordinate $l=m$ will in general cause the acceleration of a test
particle to depend to a degree (determined by the solution) on the rest mass
of the latter. \ This is a clear violation of the WEP. \ Even in other
gauges, $l$ and its associated potential $\Phi $ must be connected with the
concept of particle (as opposed to source) mass. \ We will formalize this
using the field equations below, but here we point out that such a
dependency can be expected on physical grounds: a test particle of mass $m$
in the field of a source mass $M$ only has a negligible effect on the metric
in the limit $m\diagup M\rightarrow 0$. \ The effects that follow from $%
m\diagup M\neq 0$ have traditionally been handled in areas such as
gravitational radiation by considering the ``back reaction'' of the test
particle on the field of the source [28]. \ This is clearly an approximation
to the real physics, and must break down when $m\diagup M$ is significant. \
In other words, the WEP as viewed from 5D is a geometric symmetry which must
break down at some level.

\section{\protect\underline{Vacua in 5D}}

\ \ \ \ To illustrate the argument that the 4D WEP is a symmetry of a 5D
metric, it is natural to look at solutions of the field equations that
represent a test particle in an otherwise empty space. \ Many $l$-dependent
solutions of the field equations are known, including ones for cosmology and
the solar system which are in agreement with observations [6]. \ However,
the class of solutions which represents empty 3D space has not been much
studied. \ There are technical and conceptual reasons for this. \
Technically, the field equations $R_{AB}=0$ involve in general 15 nonzero
components of the Ricci tensor. \ Even if we look for static
non-electromagnetic solutions, it is still not easy to find ones of the
desired type, which should have 3D spherical symmetry and be $\left(
r,\,l\right) $-dependent. \ Conceptually, the idea of a vacuum in 5D is
blurry. \ Even in 4D, $R_{\alpha \beta }=0$ admits solutions which are empty
of ordinary matter but have 4D curvature, the prime example being the de
Sitter solution in which spacetime is curved by the cosmological constant $%
\Lambda $, or alternatively by a vacuum fluid with density and pressure
given by $\rho _{v}=-p_{v}=\Lambda \diagup 8\pi $. \ This solution has been
extensively used in models of the origin of the classical universe based on
quantum effects, such as tunneling [23, 24]. \ In 5D, the equations $%
R_{AB}=0 $ admit solutions which are apparently empty, but whose 4D
subspaces may be curved and contain ``ordinary'' matter as determined by the
embedded Einstein equations $G_{\alpha \beta }=8\pi T_{\alpha \beta \text{ }}
$ (see above). \ A clever but only partially successful way to sidestep
these issues is to look for 5D solutions which are not only Ricci-flat with $%
R_{AB}=0$ but also Riemann-flat with $R_{ABCD}=0$ [25-27]. \ We will present
3 such solutions below, but wish to make a cautionary remark based on the
contents of the preceding section: The physical application in 4D of any $l$%
-dependent solution in 5D depends on the choice of gauge. \ The solutions
which follow are all equivalent to a flat 5D (Minkowski) manifold, but the
5D coordinate transformations which must exist between them are for
technical reasons unknown, and their different forms describe different 4D
physical vacua.

The following solutions may be confirmed by hand or computer to satisfy $%
R_{AB}=0$ and $R_{ABCD}=0$:%
\begin{eqnarray}
dS^{2} &=&\frac{l^{2}}{L^{2}}\left\{ \left( 1-\frac{r^{2}}{L^{2}}\right)
dt^{2}-\frac{dr^{2}}{\left( 1-r^{2}\diagup L^{2}\right) }-r^{2}d\Omega
^{2}\right\} -dl^{2} \\
dS^{2} &=&\frac{l^{2}}{L^{2}}\left\{ \left[ \left( 1-\frac{r^{2}}{L^{2}}%
\right) ^{1/2}+\frac{\alpha L}{l}\right] ^{2}dt^{2}-\frac{dr^{2}}{\left(
1-r^{2}\diagup L^{2}\right) }-r^{2}d\Omega ^{2}\right\}  \notag \\
&&\;\;\;\;\;\;\;\;\;\;\;\;\;\;\;\;\;\;\;\;\;\;\;\;\;\;\;\;\;\;\;\;\;\;\;\;\;%
\;\;\;\;\;\;\;\;\;\;\;\;\ \ \ \ \ \ \ \ \ \ \;\;\;\;\;\;\;-dl^{2} \\
dS^{2} &=&\frac{l^{2}}{L^{2}}\left\{ \left[ \left( 1-\frac{r^{2}}{L^{2}}%
\right) ^{1/2}+\frac{\alpha L}{l}\right] ^{2}dt^{2}-\frac{dr^{2}}{\left(
1-r^{2}\diagup L^{2}\right) }\right.  \notag \\
&&\left. \;\;\;\;\;\;\;\;\;\;\ \ \;\;\;\;\;\;\;\;\;\;-\left( 1+\frac{\beta
L^{2}}{rl}\right) ^{2}r^{2}d\Omega ^{2}\right\} -dl^{2}\;\;\;.
\end{eqnarray}%
Here $d\Omega ^{2}\equiv \left( d\theta ^{2}+\sin ^{2}\theta d\phi
^{2}\right) $, so all 3 solutions are spherically symmetric in 3D. \ The
first is a 5D canonical embedding of the 4D de Sitter solution provided the
identification $L^{2}=3\diagup \Lambda $ is made (see above). \ However, in
general $L$ measures the size of the potential well associated with $x^{4}=l$%
, as shown by the de Sitter form (12). \ Solutions like (12)-(14) depend in
general on two dimensionless constants $\alpha ,\,\beta $. \ We have
examined the properties of (12)-(14) extensively, but here note only their
generic features. \ These can be appreciated by combining (12)-(14) in the
following form:%
\begin{eqnarray}
dS^{2} &=&\frac{l^{2}}{L^{2}}\left\{
A^{2}dt^{2}-B^{2}dr^{2}-C^{2}r^{2}d\Omega ^{2}\right\} -dl^{2} \\
A &\equiv &\left( 1-\frac{r^{2}}{L^{2}}\right) ^{1/2}+\frac{\alpha L}{l}%
,\;\;B\equiv \frac{1}{\left( 1-r^{2}\diagup L^{2}\right) ^{1/2}},\;\;C\equiv
1+\frac{\beta L^{2}}{rl}\;.
\end{eqnarray}%
The 4D subspaces defined by these solutions are curved, with a 4D Ricci
scalar $^{4}R$ which by Einstein's equations is related to the trace of the
4D energy-momentum tensor by $^{4}R=-8\pi T$. \ The general expression for $%
^{4}R$ for any 5D metric of the form $dS^{2}=g_{\alpha \beta }dx^{\alpha
}dx^{\beta }+\epsilon \Phi ^{2}dl^{2}$ as used before is:

\begin{equation}
^{4}R=\frac{\epsilon }{4\Phi ^{2}}\left[ g_{\;\;,4}^{\mu \nu }\,g_{\mu \nu
,4}+\left( g^{\mu \nu }\,g_{\mu \nu ,4}\right) ^{2}\right] \;\;\;\;.
\end{equation}%
The special expression for (15), (16) is:

\begin{equation}
^{4}R=-8\pi T=-\frac{2}{L^{2}}\left[ \frac{1}{AB}+\frac{2}{ABC}+\frac{1}{%
C^{2}}+\frac{2}{C}\right] \;\;\;\;.
\end{equation}%
This shows that stress-energy is concentrated around singular shells where
one of $A$, $B$ or $C$ is zero. \ The equation of state is in general
anisotropic $\left( T_{1}^{1}\neq T_{2}^{2}\right) $. \ If one replaces $%
1\diagup L^{2}$ in (18) by its de Sitter limit $\Lambda \diagup 3$, it
becomes obvious that the meaning of \underline{the} cosmological
``constant'' requires a drastic rethink. \ The effective $\Lambda $ is in
general a function of $r$ and $l$, opening a way to a resolution of the
cosmological-constant problem. \ Indeed, there is no such thing as ``%
\underline{the} vacuum'' in 5D physics, but rather structured vacua.

\section{\protect\underline{Particle Masses in 5D}}

\ \ \ \ \ A common view, notably from inflationary quantum theory, is that
particles are intrinsically massless, gaining masses from the Higgs field
[28]. \ This view is in principle compatible with the recent demonstration
that particles which move on null paths in 5D can move on timelike paths in
4D, both for induced-matter theory [2] and brane theory [4]. \ The scalar
field $g_{44}=\epsilon \Phi ^{2}$ of 5D relativity can be suppressed by use
of one of the 5 degrees of coordinate freedom (see above); but solutions are
known for both solitons and cosmologies where $\Phi $ contains significant
physics, and it has been suggested that $\Phi $ is the classical analog of
the Higgs field [6]. \ There are in fact several ways to define the mass of
a particle in 5D. \ Here, we wish to give a short account of one which is
mathematically straightforward [16, 22] and builds on the physical
identification of the extra coordinate arrived at in section 2.

There we saw that $m=l$ for metrics of the canonical form with $\left|
g_{44}\right| =1.$ \ For metrics which are of other forms, we can define an
effective mass by%
\begin{equation}
m\equiv \int \left| \Phi \right| dl=\int \left| \Phi \left( dl\diagup
ds\right) \right| ds\;\;\;\;.
\end{equation}%
This is in line with how proper distance is defined in 3D. \ In practice, $%
\Phi $ would be given by a solution of the 5D field equations, and $%
dl\diagup ds$ would be given by a solution of the extra component of the 5D
geodesic equation (or directly from the metric for a null 5D path and a
particle at rest in 3D). \ We note that a potential problem with this
approach is that $\Phi $ may show horizon-like behaviour. \ An example is
the Gross/Perry/Davidson/Owen/Sorkin monopole, which in terms of a radial
coordinate $r$ which makes the 3D part of the metric isotropic has $%
g_{44}=-\Phi ^{2}=-\left[ \left( 1-a\diagup 2r\right) \diagup \left(
1+a\diagup 2r\right) \right] ^{2\beta \diagup \alpha }$ where $a$ is the
source strength and $\alpha $,\ $\beta $ are dimensionless constants
constrained by the field equations to obey $\alpha ^{2}=\beta ^{2}+\beta +1$
[ref. 6 p. 70]. \ This problem may be avoided by restricting the
physically-relevant size of the manifold [6, 28]. \ Another potential
problem is that real particles may have $\Phi =\Phi \left( x^{\gamma
},\,l\right) $ so complicated as to preclude finding an exact solution. \
This problem may be avoided by expanding $\Phi $ in a Fourier series:%
\begin{equation}
\Phi \left( x^{\gamma },\,l\right) =\overset{+\infty }{\underset{n=-\infty }{%
\sum }}\,\Phi ^{\left( n\right) }\left( x^{\gamma }\right) \exp \left(
i\,n\,l\diagup L\right) \;\;\;\;.
\end{equation}%
Here $L$ is the characteristic size of the extra dimension, which by (17) is
related to the radius of curvature of the embedded 4-space which the
particle inhabits. \ It should be noted that in both modern versions of 5D
relativity, namely induced-matter theory and brane theory, the extra
dimension is not compactified [1-5]. \ Thus we do not expect a simple tower
of states based on the Planck mass, but a more complicated spectrum of
masses that offers a way out of the hierarchy problem.

Underlying the comments of the preceding paragraph is the field equation $%
R_{44}=0$ which governs $\Phi $. \ The full set of field equations $R_{AB}=0$
contains 15 components. \ These can be reduced by tiresome algebra for the
general metric noted before, namely $dS^{2}=g_{\alpha \beta }\left(
x^{\gamma },\,l\right) dx^{\alpha }dx^{\beta }+\epsilon \Phi ^{2}\left(
x^{\gamma },\,l\right) dl^{2}$, which only uses 4 of the 5 degrees of
coordinate freedom to remove the potentials $\left( g_{4\alpha }\right) $ of
electromagnetic type. \ The result is sets of 10, 4 and 1 equations [6]. \
The first set comprises the Einstein equations $G_{\alpha \beta }=8\pi
\,T_{\alpha \beta }$, with $T_{\alpha \beta }$ given by (9). \ The second
set comprises the conservation equations%
\begin{eqnarray}
P_{\alpha ;\beta }^{\beta } &=&0 \\
P_{\alpha }^{\beta } &\equiv &\frac{1}{2\Phi }\left( g^{\beta \sigma
}g_{\sigma \alpha ,4}-\delta _{\alpha }^{\beta }g^{\mu \nu }g_{\mu \nu
,4}\right) \;\;\;\;.
\end{eqnarray}%
These are usually easy to satisfy in the continuous fluid of induced-matter
theory as developed by Wesson and others, and are related to the stress in
the surface $\left( l=0\right) $ of membrane theory with the $Z_{2}$
symmetry as developed by Randall and Sundrum (see ref. 5 for a discussion of
both). \ The remaining field equation is the scalar relation%
\begin{equation}
\square \Phi =-\frac{\epsilon }{2\Phi }\left[ \frac{g_{\;\;,4}^{\lambda
\beta }g_{\lambda \beta ,4}}{2}+g^{\lambda \beta }g_{\lambda \beta ,44}-%
\frac{\Phi _{,4}g^{\lambda \beta }g_{\lambda \beta ,4}}{\Phi }\right]
\;\;\;\;.
\end{equation}%
Here as before $\square \Phi \equiv g^{\alpha \beta }\Phi _{,\alpha ;\beta }$
and some of the terms on the right-hand side are present in the
energy-momentum tensor of (9). \ In fact, one can rewrite (23) for the
static case as Poisson's equation with an effective source density for the $%
\Phi $-field. \ In general, (23) is a wave equation with a source induced by
the fifth dimension. \ This supports the series expansion (20), and implies
that the inertial rest mass of a particle as defined by (19) arises from the
scalar field.

\section{\protect\underline{Conclusion}}

\ \ \ \ \ Gravity in general relativity is a force which is encoded in the
Christoffel symbols as coupled to the 4-velocities, and is inertial in the
sense that it arises from the motion of a particle with respect to a 4D
frame of reference or manifold which is not flat. \ The fifth force of
induced-matter and membrane theory is similar [1, 3]. \ The normalization
condition for the 4-velocities (1) shows that ordinary 4D geodesic motion is
augmented by a fifth force (per unit mass) or acceleration (2), which while
it depends on the velocity in the fifth dimension has the unique property of
acting \underline{parallel} to the 4-velocity (3). \ This force depends in
general on $x^{4}=l$, the fifth coordinate of the particle, and therefore
violates the Weak Equivalence Principle, at least technically. \ However, it
is compatible with the principle of conservation of linear momentum (5),
which leads to the identification of $l$ with the (inertial rest) mass of
the test particle $m$. \ Other aspects of 4D gravity support this. \ The
presence of $x^{4}=l$ in exact solutions of the 5D field equations
(12)-(14), which would otherwise be called empty, lead to the realization
that there are 5D vacua with structure. \ A definition for the rest mass $m$%
, analgous to that of proper distance and valid for any 5D metric (19), is
compatible with the identification of the scalar field of classical 5D
relativity with the Higgs field of particle physics, its field equation (23)
describing a wave with a source. \ The above conclusions clearly open ways
to resolving well-known problems that arise from mismatches of classical and
quantum physics, notably the cosmological-constant and hiearchy problems.

The WEP, however, is rendered particularly transparent. \ It is a geometric
symmetry, valid only in the limit in which the metric is independent of $%
x^{4}=l$, that is the limit where the mass of a test particle is negligible
compared to other terms such as the mass of the source. \ New techniques to
measure departures from the WEP are technically challenging [28-30]. \ But
if the 4D world is part of one with 5 or more dimensions, violations of the
WEP must exist.

\underline{{\LARGE Acknowledgements}}

This work is based on previous collaborations with H. Liu and B. Mashhoon. \
It was supported by N.S.E.R.C.

\underline{{\LARGE References}}

\begin{enumerate}
\item P.S. Wesson, B. Mashhoon, H. Liu, W.N. Sajko, Phys Lett. B \underline{%
456}, 34 (1999).

\item S.S. Seahra, P.S. Wesson, Gen. Rel. Grav. \underline{33}, 1731 (2001).

\item D. Youm, Phys. Rev. D \underline{62}, 084002 (2000).

\item D. Youm, Mod. Phys. Lett. A \underline{16}, 2731 (2001).

\item J. Ponce de Leon, Mod. Phys. Lett. A\underline{16}, 2291 (2001).

\item P.S. Wesson, Space-Time-Matter, World Scientific, Singapore (1999).

\item D.J. Gross, M.J. Perry, Nucl. Phys. B \underline{226}, 29 (1983).

\item S. Deser, M. Soldate, Nucl. Phys. B \underline{311}, 739 (1989).

\item Y.M. Cho, D.H. Park, Gen. Rel. Grav. \underline{23}, 741 (1991).

\item A. Billyard, P.S. Wesson, D. Kalligas, Int. J. Mod. Phys. D \underline{%
4}, 639 (1995).

\item P.S. Wesson, \underline{in} STEP: Testing the Equivalence Principle in
Space, ed. R. Reinhard, European Space Agency, WPP-115, 566 (1996).

\item C.M. Will, Int. J. Mod. Phys. D \underline{1}, 13 (1992).

\item J.E. Campbell, A Course of Differential Geometry, Clarendon, Oxford
(1926).

\item J.E. Lidsey, C. Romero, R. Tavakol, S. Rippl. Clkmass. Quant. Grav. 
\underline{14}, 865 (1997).

\item E. Anderson, J.E. Lidsey, Class. Quant. Grav. \underline{18}, 4831
(2001).

\item G.W. Ma, Phys. Lett. A \underline{143}, 183 (1990).

\item B. Mashhoon, H. Liu, P.S. Wesson, Phys. Lett. B \underline{331}, 305
(1994).

\item P.S. Wesson, H. Liu, Int. J. Theor. Phys. \underline{36}, 1865 (1997).

\item B. Mashhoon, P.S. Wesson, H. Liu, Gen. Rel. Grav. \underline{30}, 555
(1998).

\item W.B. Belayev, gr-qc/0110099 (2001).

\item J. Ponce de Leon, Phys. Lett. B \underline{523}, 311 (2001).

\item P.S. Wesson, J. Math Phys. in press (2002) (gr-qc/0105059).

\item A. Vilenkin, Phys. Lett. B \underline{117}, 25 (1982).

\item A. Vilenkin, Phys. Rev. D \underline{37}, 888 (1988).

\item G. Abolghasem, A.A. Coley, D.J. McManus, J. Math Phys. \underline{37},
361 (1996).

\item H. Liu, P.S. Wesson, Gen. Rel. Grav. \underline{30}, 509 (1998).

\item P.S. Wesson, H. Liu, Phys. Lett. B \underline{432}, 266 (1998).

\item R.B. Mann, P.S. Wesson (eds.), Gravitation: A Banff Summer Institute,
World Scientific, Singapore (1991).

\item R. Reinhard (ed.), STEP: Testing the Equivalence Principle in Space,
European Space Agency, WPP-115 (1996).

\item R.T. Jantzen, G.M. Keiser, R. Ruffini (eds.), Proc. 7th. Marcel
Grossman Meeting, World Scientific, Singapore (1996).
\end{enumerate}

\end{document}